\documentstyle[12pt,epsf]{article}

\newcommand{\vspu}{\vspace{5mm}}

\newcommand\be{\begin{equation}}
\newcommand\ee{\end{equation}}
\newcommand\bea{\begin{eqnarray}}
\newcommand\eea{\end{eqnarray}}

\makeatletter
\@addtoreset{equation}{section}
\makeatother
\begin{document}
\baselineskip=15pt

\begin{center}
{\large\bf Ginsparg-Wilson Fermions: \\
A study in the Schwinger Model
\footnote{This work is supported in part by funds provided by the U.S.
Department of Energy (D.O.E.) under cooperative research agreement
DE-FG02-96ER40945.}}
\vspu \\
Shailesh Chandrasekharan\footnote{ email:sch@phy.duke.edu}
\vspu \\

Department of Physics, \\
Box 90305, Duke University \\
Durham NC 27708-0305, USA \\
\vspu

Preprint: \ DUKE-TH-98-177 \\
\vspu
\end{center}
\vspu

\begin{abstract} 
\normalsize
Qualitative features of Ginsparg-Wilson fermions, as formulated by
Neuberger, coupled to two dimensional $U(1)$ gauge theory are studied. The
role of the Wilson mass parameter in changing the number of massless
flavors in the theory and its connection with the index of the Dirac 
operator is studied. Although the index of the Dirac operator is not 
related to the geometric definition of the topological charge for strong 
couplings, the two start to agree as soon as one goes to moderately weak 
couplings. This produces the desired singularity in the quenched chiral 
condensate which appears to be very difficult to reproduce with staggered 
fermions. The fermion determinant removes the singularity and reproduces 
the known chiral condensate and the meson mass within understandable errors. 
\end{abstract}

\newpage

\begin{center}
\noindent{\bf 1. Introduction}
\end{center}

  Our understanding of chiral symmetry on the lattice has matured
recently. The overlap formulation \cite{Narayanan} has lead to a new 
Dirac operator that is expected to describe exactly massless 
fermions \cite{Neuberger1}. The chiral symmetry on the lattice that 
makes this possible has been discovered in Ref. \cite{Luscher}. This symmetry 
arises when the Dirac operator $D$ satisfies\footnote{The 
relation is more general. We only state the simplest form here.}
\begin{equation}
\label{eq:GWR}
D\gamma_5 + \gamma_5 D = D \gamma_5 D,
\end{equation}
called the Ginsparg-Wilson relation \cite{Ginsparg}. This relation 
appears to be quite powerful. It helps to reproduce the continuum relations 
arising due to chiral symmetry, and thus leads to elegant renormalization 
properties \cite{Hasenfratz4}. Further, like in the continuum, it also 
suggests an elegant resolution of the $U(1)$ problem on the lattice which has 
interesting consequences for the lattice calculations of QCD \cite{Chandra}. 
We will refer to the fermion formulations which obey the Ginsparg-Wilson 
relation as Ginsparg-Wilson fermions. 

In addition to the Neuberger operator \cite{Neuberger2}, the fixed point 
Dirac operator also obeys the Ginsparg-Wilson relation \cite{Hasenfratz2}.
Here we do not discuss the latter approach. In the context of the Neuberger's
Dirac operator it is important to point out that  there is another closely 
related approach to massless fermion based on an extension \cite{Shamir} of 
an idea of Kaplan \cite{Kaplan} called domain wall fermions. These fermions
can be considered as a truncated version of the overlap \cite{Neuberger3}. 
The appropriate Dirac operator of domain wall fermions satisfy the 
Ginsparg-Wilson relation up to exponentially small violations in the 
limit where the separation of the walls, on which the physical fermions live, 
becomes large. The domain wall
fermions have been used in QCD calculations \cite{Blum} and appear to have 
very good chiral properties compared to the Wilson fermion approach. 
More recently this approach has also been shown to reproduce some of the 
singularities in the chiral condensate due to exact zero modes that have 
not been seen in staggered fermion formulations \cite{Columbia}. 
Thus the domain wall fermions appear to be a practical way to implement 
the Ginsparg-Wilson relation approximately. However, in this article we 
explore the possibility of keeping the Ginsparg-Wilson relation exact by 
working directly with Neuberger's Dirac operator.

Although chiral symmetry emerges naturally with Ginsparg-Wilson fermions, 
their dynamical consequences are just beginning to be explored. The first 
study of the fixed point Dirac operator was done in the Schwinger 
model in Ref. \cite{Farchioni}. A similar study for Neuberger's Dirac 
operator is 
lacking\footnote{Look at the work by Farchioni et al., \cite{Farchioni2},
which appeared after the completion of our present work.}. Tests performed
in closely related formulations like the overlap \cite{OverlapTests} 
and domain wall fermions \cite{Pavlos98} can only be used to gain confidence, 
since all these formulations differ in their cutoff effects. In Neuberger's
Dirac operator such effects enter through a mass parameter $M$. It is 
important to understand such effects and check for universality. This has 
not yet been done. Further, in the context of the Schwinger model, the
studies in the overlap and domain wall formulations have only concentrated
on the chiral condensate \cite{Pavlos,Thirring}\footnote{ A 
rather thorough analysis of the chiral condensate for the two flavor theory 
in the domain wall formalism has been done in Ref. \cite{Pavlos98}.}. The
meson mass has been ignored. As we discuss below, this is another
important quantity to calculate to confirm universality.

A new feature of the Ginsparg-Wilson fermions is that the Dirac 
operator has a non-zero coupling between any two lattice points which are 
exponentially small in the distance between the points \cite{Horvath98}.
Such actions are still considered local\footnote{Actions that have only 
finite range interactions are referred to as ultra-local.} if they do not 
destroy the mass spectrum of the theory. This can be checked in the free 
theory. On the other hand it is important to see if this feature continues 
at a non-perturbative level and no spurious long range correlations 
originate. This question has been recently addressed for 
smooth gauge fields in Ref. \cite{Hernandez98}. Non-perturbatively one 
can study the mass spectrum of the theory. Thus determining the meson mass 
in the Schwinger model is important to confirm our beliefs.

In this article we study a two dimensional $U(1)$ lattice gauge theory 
coupled to fermions proposed by Neuberger \cite{Neuberger1}. In the continuum 
limit, this model approaches the Schwinger model up to four-Fermi 
couplings \cite{Thirring}\footnote{I thank Pavlos Vranas for reminding me 
of this fact.}. To control these couplings is quite difficult.
and will not be attempted here. Instead we will try to learn qualitative
features of the model as a way to learn how it could reproduce the main
properties of the Schwinger model if a more thorough and complete analysis 
is made. We also leave the more interesting multi- flavor case to later 
studies since there the dynamics is considerably more complicated and 
finite volume effects must be carefully studied before the effects of 
complete chiral symmetry guaranteed by Neuberger's Dirac operator can be 
understood. In particular it is important to obtain results like 
$\langle\bar\psi\psi\rangle \sim m^{1/3}$ and the pion mass
$\sim m^{2/3}$ \cite{Coleman76,Pavlos98}.

This article is organized as follows. In section 2 we introduce the model.
In section 3 we discuss the role of the Wilson mass parameter that enters 
the definition of the Dirac operator. In particular we show an interesting
relation between the zero modes of the Dirac operator and the number of 
fermion flavors that the theory describes for background gauge field 
configurations. In section 4 we discuss the connection between 
topology and index of the Dirac operator. This was recently studied for 
semi-classical gauge field configurations in Ref. \cite{Chiu}. Here we 
understand this
connection as a function of the gauge coupling. We then argue that this is 
responsible for reproducing the singularities of the quenched approximation. 
In section 5 we calculate the spectral density of the 
Dirac operator as a function of the fermion mass. We also calculate the chiral 
condensate and the meson mass after including the fermion determinant and 
compare it with the results in the Schwinger model. A short discussion
showing the relevance of the results to QCD and directions for future work 
are pointed out in section 6.

\begin{center}
\noindent{\bf 2. The Model}
\end{center}

  The model that we investigate can be described by the action
\begin{equation}
\label{gact}
S = \beta\sum_{\cal P}\left( 1 - \frac{1}{2}
[U_{\cal P} + U_{\cal P}^\dagger] \right) + 
\sum_{x,y} \bar\Psi_x [D_{x,y} + m\delta_{x,y}] \Psi_y
\end{equation}
where $U_{\cal P}$ is the phase obtained by multiplying the four $U(1)$
phase angles attached to a plaquette and $\beta = 1/g^2$ is the gauge 
coupling. Fermions of mass $m$ interact with 
the gauge fields through the Dirac operator $D$, chosen to be the one 
proposed by Neuberger \cite{Neuberger1}. In terms of the hermitian 
Wilson-Dirac operator 
\begin{eqnarray}
H_{x,y} &=& -\frac{\gamma_5}{2}\left(
  \delta_{x,y-\hat\mu}U^\dagger_\mu(x)
\left[1 - \gamma_\mu\right]
+ \delta_{x,y+\hat\mu}U_\mu(y)
\left[1 + \gamma_\mu\right] \right) 
+ \delta_{x,y} \gamma_5(2+M),
\end{eqnarray}
with the mass parameter $M$, $D$ is given by
\begin{equation}
\label{eq:Dirac}
D = 1 + \gamma_5 H/\sqrt{H^\dagger H}.
\end{equation}
It is easy to show that $D$ satisfies eq.(\ref{eq:GWR}). The fermions obey
(anti)periodic boundary conditions in the (temporal) spatial directions
and the gauge fields are periodic.

 The condensate that characterizes chiral symmetry breaking was defined 
in Ref. \cite{Chandra}. After integrating out the fermions this can be 
written as
\begin{equation}
\label{eq:psi}
\langle\bar\psi\psi\rangle = 
Z\frac{1}{L_s L_t}\left\langle{\rm Tr} \left[(1 - \frac{D}{2})\frac{1}{D+m} 
[{\rm Det}(D+m)]^{N_f}\right] \right\rangle
/\langle [{\rm Det}(D+m)]^{N_f} \rangle
\end{equation}
where $L_s,L_t$ are the lattice sizes in the spatial and temporal
directions and $Z$ takes care of the wave function normalization. $N_f$
is the number of flavors. $N_f=0$ corresponds to the quenched theory. We 
do not consider $N_f \geq 2$ in this article.

All our calculations are performed on an $8\times 16$ lattice. Gauge fields 
are generated by a combination of head bath and over relaxation algorithm. 
Every gauge field configuration is generated after 500 sweeps 
(one sweep = one heat bath update + 5 over-relaxation step over the entire 
lattice). The fermion determinants are absorbed into the observable and
the expectation value is taken over gauge field configurations generated 
with the Boltzmann weight of the pure gauge theory. Although not a very 
reliable method for large lattices, the procedure seems to produce results 
with controlled errors in the present calculation. This is because the
physics of the one flavor model is mainly sensitive to topology and the 
fermion determinant within a topological sector does not fluctuate much. 
Further these sectors are updated well with the above algorithm on lattice 
sizes such as ours. Calculations involve a statistics over 2500 independent
gauge field configurations. The errors are obtained through a jack-knife
procedure.

The model contains three additional parameters. The coupling 
$g=1/\sqrt{\beta}$ controls the lattice spacing. Since $g$ is dimensionful 
in two dimensions, $g \sim 1/a$. The continuum limit of this model is 
reached in the $\beta\rightarrow \infty$ limit. The mass parameter $M$ 
does not effect the fermion mass, which is determined by $m$. It controls the
effects of the fermion doubling. We will clarify its role further in the 
next section. Since it is difficult to control the four Fermi-couplings 
induced by the above model, it is considerably more difficult to do a 
systematic continuum extrapolation. Based on results in Ref. 
\cite{Thirring} it 
is likely that in the continuum limit the above model will show differences 
with the results in the Schwinger model at the $10\%$ level. Hence we study 
the model at a fixed $\beta$, which also introduces finite lattice spacing 
errors. As we will see the results from the present study differ from the 
Schwinger model by $10\%-15\%$, and based on the above discussion can be
considered understandable.

\begin{center}
\noindent{\bf 3. The role of $M$}
\end{center}

The parameter $M$ is the mass of the Wilson Dirac operator that enters 
the definition of $D$. The theory can describe massless fermions for
a range of values of $M$. Interestingly, as $M$ changes it can change 
the number of massless fermions in the theory and thus can reintroduce 
fermion doubling. This happens at some critical values, say $M_c$. It is
easy to understand this in the free theory by analyzing the propagator. 
The free propagator at $M = -1$ was constructed in 
Ref. \cite{Chiu}\footnote{The free propagator for a whole class of 
Ginsparg-Wilson fermions was discussed in Ref. \cite{Bie98}.}.
Extending this result to arbitrary $M$ we get
\begin{equation}
S_F(p) = \frac{1}{2} - \frac{i\gamma_\mu \sin(p_\mu)}{2[N(p)+u(p)]}
\end{equation}
where $-\pi < p \leq \pi$, with
\begin{equation}
u(p) = 2 + M - \sum_\mu \cos(p_\mu),\;\;\; 
N(p) = \sqrt{u^2(p) + \sum_\mu \sin^2(p_\mu)}
\end{equation}
When $M>0$ there are no massless particles. For $-2 < M < 0$ the only 
pole in the propagator describing massless particles is at the origin of the 
$(p_1,p_2)$ plane. However, in the region $-4 < M < -2$  two additional 
poles appear at $(\pi,0)$ and $(0,\pi)$. Here the Dirac operator describes
three flavors of massless fermions. Finally for $M < -4$ a fourth pole at 
$(\pi,\pi)$ also appears and the theory describes four massless fermions. 
Thus the theory can describe different phases which are distinguished by
the number of massless fermion flavors. In the free theory the critical
values $M_c=0,-2$ and $-4$ separate these phases. The number of massless 
flavors in the various phases discussed here differ from the those in
domain wall fermions when the extra dimension is discrete \cite{Karl}. 
This question was recently investigated by Shamir who observes the above 
results in the limit where the lattice spacing in the extra dimension 
becomes zero \cite{Shamirnew}.

The critical values of $M_c$ will fluctuate in the presence of gauge fields. 
It is important to find a region in $M$ in the full interacting theory 
where one can isolate the one flavor phase. In the weak coupling 
limit such a phase is most likely to exist. Since this also coincides 
with the continuum limit, both for the present model and in QCD, it is 
likely that one can actually obtain the physics of one flavor, sufficiently 
close to the continuum. On the other hand in realistic calculations 
it will be important to establish this before the results can be trusted. 
We will show that there is an interesting connection between the index (or
equivalently the number of zero modes with a given chirality) and the 
number of massless fermions described by the Dirac operator in
a background gauge field configuration at a given value of $M$. It may
be possible to use this connection to determine the values of $M$ that belong
to the one flavor phase for each given background gauge field configuration.

Consider a background gauge field configuration with a non-trivial
index in the one flavor phase. As we cross a critical value of $M$ the
index can change. Such effects have been studied recently in 
Ref. \cite{Edwards}.
The change appears to be related to the change in the number of massless 
flavors the Dirac operator describes. We can predict it as 
a function of $M$ based on this relationship. For example, let $\nu\neq 0$ 
be the index of $D$ in the one flavor phase. In the three flavor phase, we 
expect massless fermions from the 
poles at $(0,0)$, $(\pi,0)$ and $(0,\pi)$. However, the chirality 
definition of the fermions with poles at $(\pi,0)$ and $(0,\pi)$ are
opposite to that of $(0,0)$ and $(\pi,\pi)$. This means that in the three
flavor phase the index of $D$ will be $-\nu$. A similar argument shows that
the index of the operator in the four flavor case would be zero. Since the
operator is designed to produce zero modes equal to the index, this means
that the Dirac operator will not have any (except accidental) zero modes in 
the zero and four flavor phases and will have $\nu$ zero modes of opposite
chirality in the one and three flavor phases. Thus one can scan through $M$ 
to find the four phases for any gauge field with a non-zero index. 
In order to check the above observations numerically, we generated a 
series of gauge field configurations at $\beta=3.$ on a $8\times 16$ lattice. 
In table \ref{tab:top.vs.M} we give values of the index of the Dirac 
operator for ten such configurations for four values of $M$ in the various 
phases. The results indicate that $M=-1.0$ is a region dominated by the 
physics of one fermion flavor, when $\beta \geq 3.$ 

\begin{table}[htb]
\begin{center}
\begin{tabular}{|c|c|c|c|c|}
\hline
\multicolumn{1}{|c}{ } & \multicolumn{4}{|c|}{Index} \\ \hline
Gauge Action & M=1.0 & M=-1.0 & M = -3.0 & M = -5.0 \\ \hline
 0.1665 &  0  & -1 &      1  &  0  \\
 0.1589 &  0  &  1 &     -1  &  0  \\
 0.1962 &  0  &  3 &     -3  &  0  \\
 0.1669 &  0  &  0 &      0  &  0  \\
 0.1822 &  0  &  0 &      0  &  0  \\
 0.2309 &  0  & -1 &      1  &  0  \\
 0.1933 &  0  &  2 &     -2  &  0  \\
 0.1904 &  0  &  0 &      0  &  0  \\
 0.1753 &  0  &  1 &     -1  &  0  \\
 0.1527 &  0  &  0 &      0  &  0  \\ \hline
\end{tabular}
\end{center}
\label{tab:top.vs.M}
\caption{\em The table shows the index of the Dirac operator as a 
function of the Wilson mass $M$. The connection of the index with
the number of massless fermion flavors and their chiralities is 
responsible for the above results. The first column indicates the
value of the plaquette term of eq.(\ref{gact}) for the various 
configurations used. The results are consistent with theoretical
predictions.
}
\end{table}

As far as we know, the above connection between the index of the Dirac 
operator and the number of fermion flavors has not been appreciated before. 
In fact this connection leads to an amusing conjecture. In the region 
$M<-4$ there are four massless flavors but the Dirac operator does not 
produce any zero modes related to topology. The zero modes, if they exist, 
are accidental. However, there is a $U(1)$ chiral symmetry in the action 
because the operator obeys the Ginsparg-Wilson relation \cite{Luscher}. 
Since there are no zero modes to break the symmetry through the measure, 
the exact chiral symmetry present must be flavored and hence not anomalous. 
This reminds one of staggered fermions where such a symmetry exists 
although the theory describes two flavors of massless 
fermions .

In addition to producing the different phases, $M$ is expected to 
renormalize the relevant couplings including the gauge coupling and the
fermion mass. In fact it also introduces a wave-function renormalization. This
can be seen by expanding the fermion propagator in small momenta near 
the poles in the various phases. We get
\begin{equation}
2[N(p)+u(p)] = \left\{ 
 	\begin{array}{cc} 
	\frac{1}{|M|} (p_1^2 + p_2^2) + ...
			&  \hbox{expansion near}\; (0,0),\;\; M < 0\\
	\frac{1}{|2+M|} (p_1^2 + p_2^2) + ...
			&  \hbox{expansion near}\; 
(0,\pi)\;\; {\rm or}\;\; (\pi,0),
				\;\;M < -2 \\
	\frac{1}{|4+M|} (p_1^2 + p_2^2) + ...
			&  \hbox{expansion near}\; (\pi,\pi),\;\;\;
				M < -4 \\
	\end{array}
	\right.
\end{equation}
This shows that the wave-function renormalization at tree level turns out to
be $Z=1/|M|$ in the one flavor phase. As we will see below, the dominant
effect of $M$ on the chiral condensate of eq.(\ref{eq:psi}) comes
from $Z$. Interestingly, we see that the residue at the various poles are 
different showing that the various massless fermions can have different
wavefuntion renormalization constants. This suggests that the 
renormalization properties of the three and four flavor phases needs 
to be understood further before they can be treated as a regularized 
version of legitimate field theories.

\begin{center}
\noindent{\bf 4. Zero Modes and Quenching}
\end{center}

\begin{figure}[htb]
\hskip0.7in
\epsfysize=90mm
\epsffile{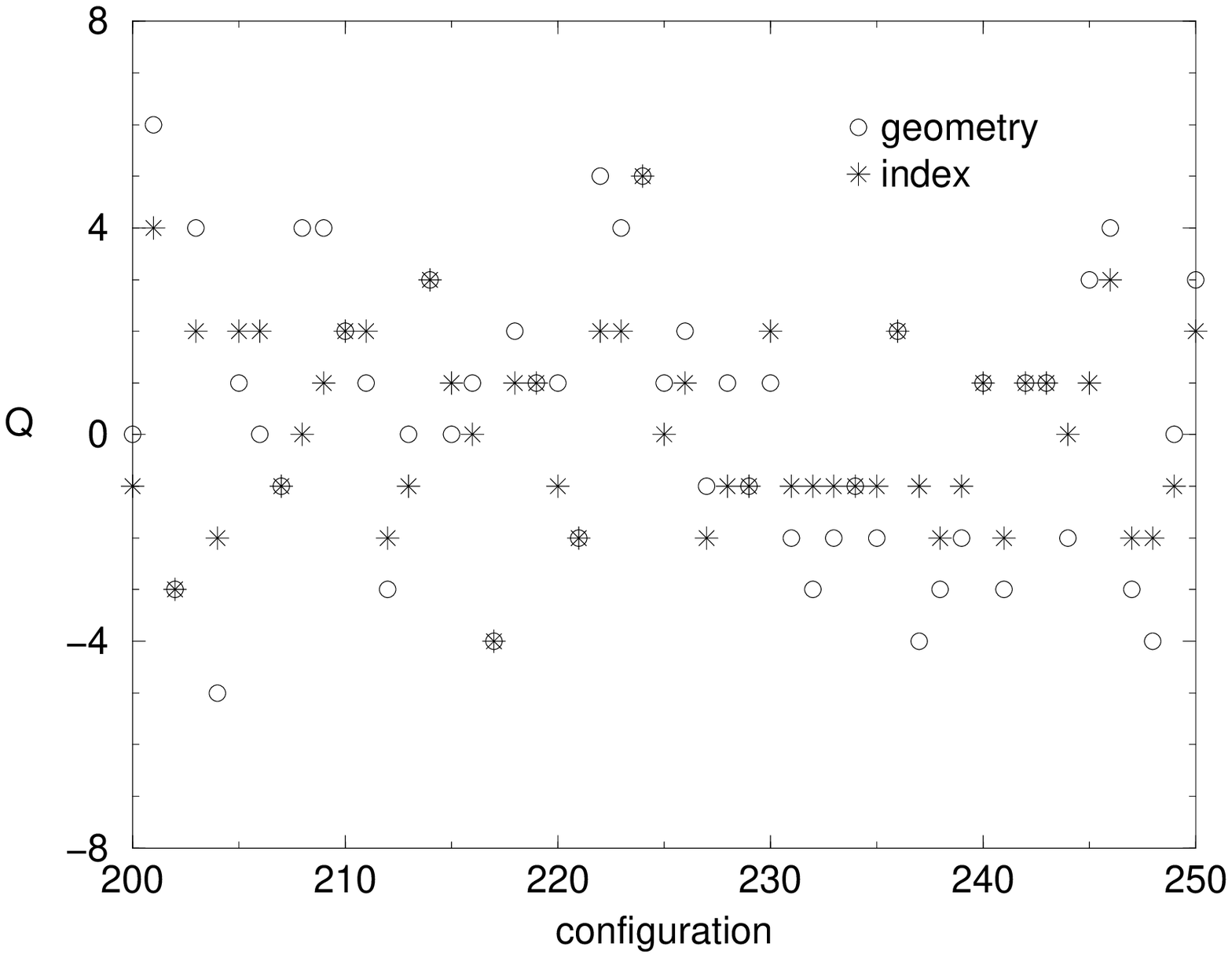}
\vskip0.5in
\hskip0.7in
\epsfysize=90mm
\epsffile{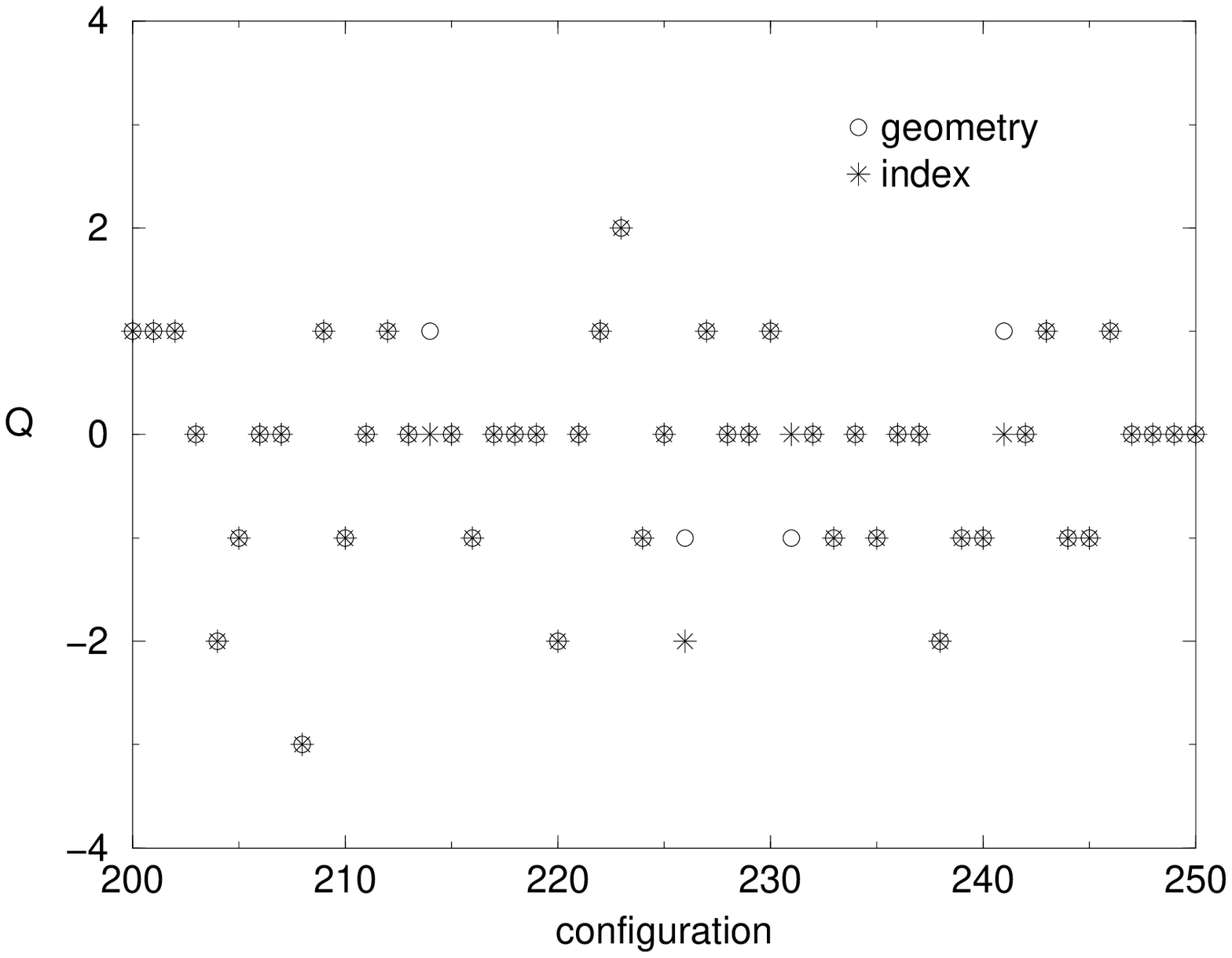}
\caption{\it The geometric definition of the topological charge is
compared with the index of the Dirac operator for 50 configurations 
generated at $\beta=0.5$ (top) $\beta=3.0$ (bottom).}
\label{fig:top.vs.geom}
\end{figure}

\begin{figure}[hbt]
\hskip1in
\epsfysize=80mm
\epsffile{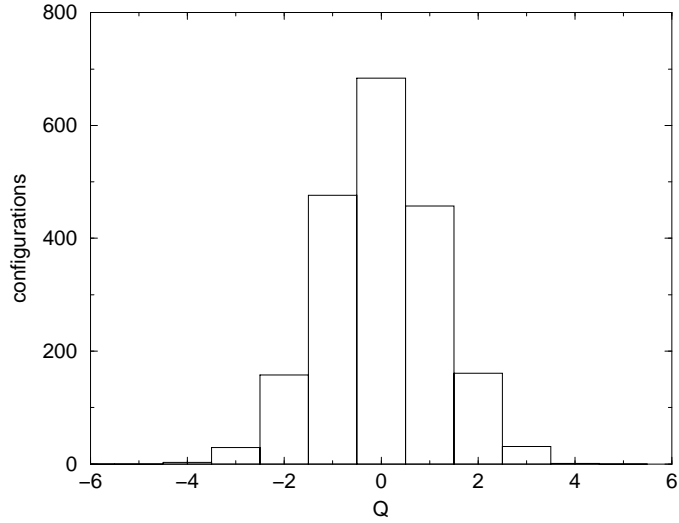}
\caption{\it This figure shows a histogram of the topological sectors
generated at $\beta=3.0$ in the 2500 configurations that were analyzed.}
\label{fig:hist}
\end{figure}

The one flavor Schwinger model is mainly governed by the dynamics of the
gauge field topology and its connection with zero modes of the Dirac 
operator through the index theorem. Previous studies have shown that
Ginsparg-Wilson fermions can reproduce the index theorem easily. We 
confirm this here. We choose the geometric definition of the topological 
charge given by
\begin{equation}
Q = \frac{1}{2\pi} \sum_{\cal P} \log(U_{\cal P})
\end{equation}
where the value of the log is defined on the principal sheet, i.e., 
$0 \leq \log(U_{\cal P}) < 2\pi$. For smooth fields one expects that the 
index of the Dirac operator will be related to the above definition of 
the topological charge \cite{Chiu}. For rough fields there is no reason 
for such a relation. In order to quantify this effect we compared Q with 
the index of the Dirac operator at a strong coupling $\beta=0.5$ and a 
moderately weak coupling $\beta=3.0$ lattice. The results for fifty 
configurations are plotted in the figure (\ref{fig:top.vs.geom}).
These results show that 
even though the index of the Dirac operator is not related to the 
geometrical definition in any formal sense, the two start to agree quite 
well at moderately weak couplings and that $\beta=3$ is sufficiently weak
to reproduce the topological properties of the two dimensional theory.
In figure (\ref{fig:hist}) we plot the histogram for the topological sectors 
obtained from an ensemble of 2500 configurations at $\beta=3$. This 
further shows that the algorithm samples the various topological sectors 
adequately.

The realization of the index theorem on the lattice has important 
consequences. In QCD it is well known that the mass of the $\eta^\prime$
particle is a consequence of the dynamics of the zero modes of the Dirac 
operator. In the quenched theory the zero modes are enhanced since 
the suppression of these modes due to the fermion determinant is absent. 
This causes many interesting singularities in the theory. In the
infinite volume limit, these singularities have been predicted based on
quenched chiral perturbation theory in Ref. \cite{Sharpe}. At 
small volumes
semi-classical expansions suggest that there will be instanton like
configurations that will dominate the path integral and this will lead to 
exact zero modes of the Dirac operator. This leads to new singularities.
One of the consequences is the divergence of the quenched chiral 
condensate. 

\begin{figure}[hbt]
\hskip1in
\epsfysize=80mm
\epsffile{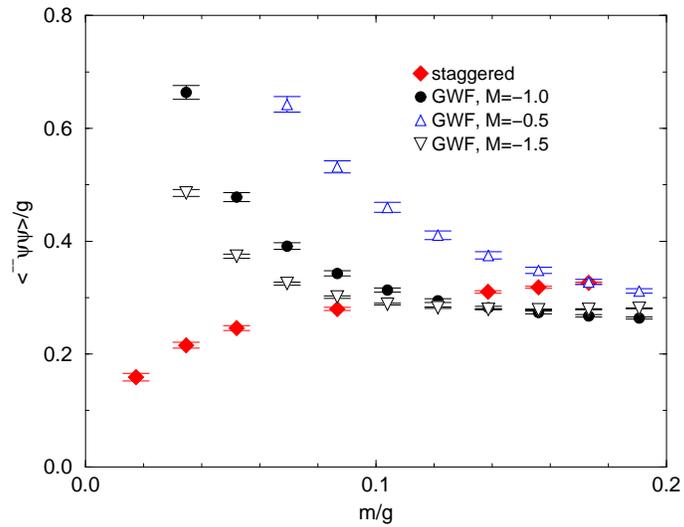}
\caption{\it The chiral condensate including the factor $Z = 1/|M|$
in the quenched limit is shown. For comparison we also include the 
results from staggered fermion simulations.}
\label{fig:qpsi}
\end{figure}

The divergence of the quenched chiral condensate in the Schwinger model
has been discussed in Ref. \cite{Doel84}. The condensate was first calculated 
in Ref. \cite{Marinari} using staggered fermions on a $16\times 16$ lattice.
The divergence as a function of the fermion mass was not found and was 
completely ignored even in the discussion. The singularity was first 
discussed in some detail in Ref. \cite{Kenway} 
where simulations on $64\times 64$ lattices were reported. 
These calculations have not been confirmed by other groups. Recent studies 
in QCD suggest that such singularities are difficult to obtain with 
staggered fermions \cite{Kaehler,Chandra96}, but appear rather easily in the 
domain wall formalism \cite{Columbia}. The above discussion of the index theorem
shows that with Ginsparg-Wilson fermions these divergences must appear 
easily. In order to show the dramatic difference between staggered fermions 
and the Ginsparg-Wilson fermions, we plot the quenched chiral condensate as 
a function of the fermion mass in figure (\ref{fig:qpsi}) for both types 
fermions.

\begin{center}
\noindent{\bf 5. Dirac Spectral Density and Meson Mass }
\end{center}

  The study of the chiral condensate and the meson mass is
important in any study of the Schwinger model. The chiral condensate 
is given by eq.(\ref{eq:psi}). Written in terms of the 
eigenvalue density $\rho(\theta)$ of the Dirac operator \cite{Chandra},
it reads 
\begin{equation}
\langle\bar\psi\psi\rangle = Z \int_{-\pi}^\pi d\theta\;\rho(\theta) 
\frac{1 + \cos(\theta) + i\sin(\theta)}
{2(1 - \cos(\theta) - i\sin(\theta) + am)}.
\end{equation}
In deriving the above representation, we have used the fact that the 
eigenvalues of the Ginsparg-Wilson Dirac operator\footnote{We assume that
$D$ obeys $D^\dagger = \gamma_5 D \gamma_5$.} are distributed on a 
unit circle in the complex plane centered at $(1,0)$. The density is 
normalized using the relation $\int d\theta \rho(\theta) = 2$.

\begin{figure}[hbt]
\hskip1in
\epsfysize=80mm
\epsffile{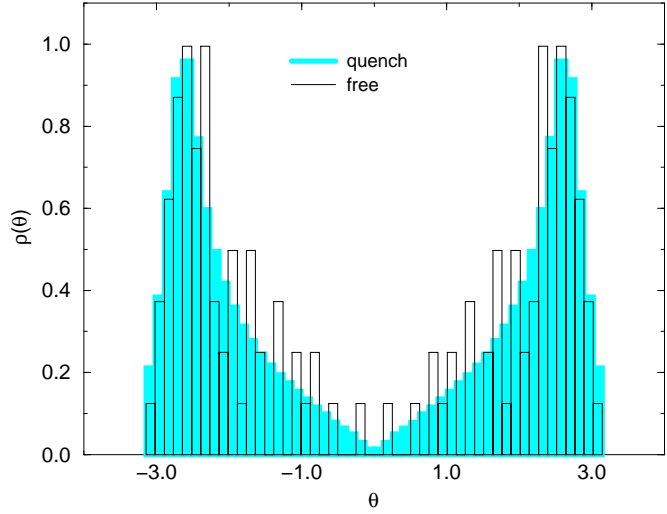}
\caption{\it This figure shows the eigenvalue density of $D$ for the
quenched run at $\beta=3.0$ and for free field theory.}
\label{fig:rho}
\end{figure}

An eigenvalue at $\theta=0$ is an exact zero mode. It is easy to see that
every eigenvalue at $\theta=0$ will be accompanied by an eigenvalue at
$\theta=\pi$ for topological reasons. Separating these topological zero 
modes, we find
\begin{equation}
\rho(\theta) = \rho_0\left[\delta(\theta) + \delta(\theta-\pi)\right]
+ \rho^\prime(\theta)
\end{equation}
In figure (\ref{fig:rho}) we plot $\rho^\prime(\theta)$ at $\beta=3.0$ and 
$|M|=1.0$ obtained using quenched configurations. For comparison we 
also plot the free field eigenvalue density.

It is useful to understand the dependence of $\rho(\theta)$ on the
dynamical fermion mass. Since all the eigenvalues are measured, this is 
easily obtained. For each configuration we divide the
range $(-\pi,\pi)$ into 50 bins and count the number of eigenvalues 
falling into each bin and divide it by $(L_s L_t)2\pi/50$ to obtain
the value of $\rho^\prime(\theta)$ where $\theta$ refers to the 
center of the bin. For $\theta = 0$ we give the value of $\rho_0$
which is obtained by counting the zero modes and dividing by the 
$L_s L_t$. $\rho(\theta)$ for four values of the dynamical fermion mass
is given in table \ref{tab:rhodyn}. The errors are obtained by a 
jackknife method. $\rho(\theta)$ is even and we give the values for
positive $\theta$. From the results it appears that the statistics of 
2500 configurations is far from sufficient to see the effects of the 
fermion determinant except in the region $|\theta| < 0.35$. This
corresponds to masses less than 0.06 in lattice units.

\begin{figure}[hbt]
\hskip1in
\epsfysize=80mm
\epsffile{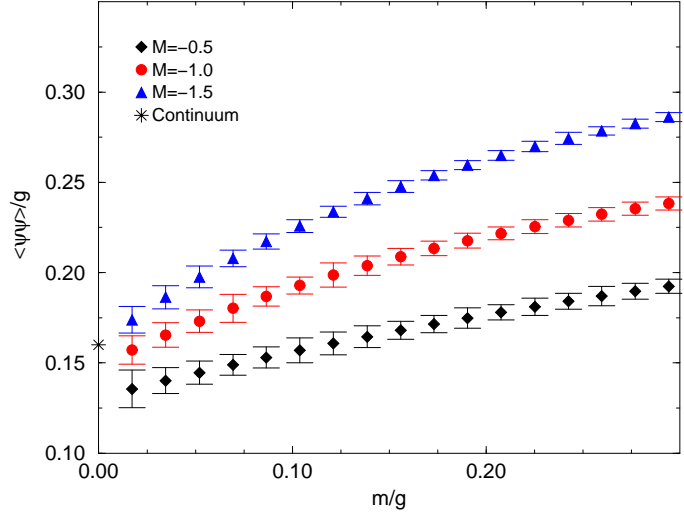}
\caption{\it The chiral condensate including the factor $Z = 1/|M|$
is plotted as a function of the fermion mass for various values of $M$.
The dominant role of $|M|$ comes through $Z$. The continuum value in the
massless Schwinger model is shown on the $m/g=0$ axis.}
\label{fig:psi}
\end{figure}

The chiral condensate is evaluated for three different values of $M$, in 
the one flavor phase, in order to study its effects. The dominant effect 
comes from the tree level wave function renormalization factor $Z=1/|M|$. 
This factor will get corrections from loop effects. The results after 
including this factor are shown as a function of the fermion mass in figure 
(\ref{fig:psi}). In the continuum it can be shown that
\begin{equation}
\label{eq:exact}
\frac{\langle\bar\psi\psi\rangle}{g} = 
\frac{{\hbox {\rm \bf C}}}{2\pi\sqrt{\pi}},
\end{equation}
where ${\hbox {\rm \bf C}}=1.78107...$ is the Euler's constant. 
This is indicated on the $m=0$ axis. As discussed in section 2, it 
will be difficult to reproduce the results of the Schwinger model with 
errors of less than $10-15\%$ from the present lattice study. Our
results are consistent with this expectation.

\begin{table}[htb]
\begin{center}
\begin{tabular}{|ccccc|}
\hline
\multicolumn{1}{|c|}{$\theta$} & \multicolumn{4}{c|}{$\rho(\theta)$} \\
\hline
  & m=0.01  & m = 0.1   &  m = 1.0  &  quenched \\
\hline
 3.0788  &  0.180(08)  & 0.189(04)  &  0.209(01)   & 0.2110(12)   \\
 2.9531  &  0.411(22)  & 0.395(09)  &  0.384(02)   & 0.3839(10)   \\
 2.8274  &  0.655(38)  & 0.648(17)  &  0.638(03)   & 0.6379(11)   \\
 2.7018  &  0.926(60)  & 0.922(33)  &  0.914(03)   & 0.9130(09)   \\
 2.5761  &  0.978(52)  & 0.970(23)  &  0.960(03)   & 0.9590(11)   \\
 2.4504  &  0.755(44)  & 0.762(18)  &  0.770(02)   & 0.7697(09)   \\
 2.3248  &  0.573(34)  & 0.583(22)  &  0.596(03)   & 0.5965(16)   \\
 2.1991  &  0.503(28)  & 0.498(12)  &  0.495(02)   & 0.4948(10)   \\
 2.0735  &  0.431(25)  & 0.425(11)  &  0.417(02)   & 0.4176(11)   \\
 1.9478  &  0.370(22)  & 0.364(09)  &  0.359(02)   & 0.3590(12)   \\
 1.8221  &  0.310(17)  & 0.310(07)  &  0.312(02)   & 0.3130(13)   \\
 1.6965  &  0.264(15)  & 0.270(07)  &  0.277(02)   & 0.2774(14)   \\
 1.5708  &  0.236(15)  & 0.240(06)  &  0.245(01)   & 0.2448(10)   \\
 1.4451  &  0.219(13)  & 0.219(06)  &  0.217(02)   & 0.2174(09)   \\
 1.3195  &  0.203(13)  & 0.199(06)  &  0.194(01)   & 0.1938(09)   \\
 1.1938  &  0.185(12)  & 0.180(05)  &  0.175(01)   & 0.1745(07)   \\
 1.0681  &  0.151(10)  & 0.152(04)  &  0.154(01)   & 0.1537(09)   \\
 0.9425  &  0.127(09)  & 0.131(07)  &  0.135(01)   & 0.1350(09)   \\
 0.8168  &  0.109(08)  & 0.112(03)  &  0.116(01)   & 0.1161(09)   \\
 0.6912  &  0.099(05)  & 0.098(03)  &  0.099(01)   & 0.0988(08)   \\
 0.5655  &  0.083(06)  & 0.082(03)  &  0.079(01)   & 0.0784(08)   \\
 0.4398  &  0.070(04)  & 0.068(02)  &  0.0648(7)   & 0.0645(08)   \\
 0.3142  &  0.069(04)  & 0.059(02)  &  0.0502(7)   & 0.0500(08)   \\
 0.1885  &  0.036(03)  & 0.029(01)  &  0.0286(6)   & 0.0289(07)   \\
 0.0628  &  0.008(01)  & 0.008(01)  &  0.0129(5)   & 0.0131(05)   \\
 0.0000  &  0.0008(1)  & 0.0044(1)  &  0.0068(2)   & 0.0070(02)   \\
\hline
\end{tabular}
\end{center}
\caption{ The table gives the eigenvalue density of the Dirac operator
as a function of the fermion mass that appears in the determinant. The
last column refers to the infinite mass limit (which is equivalent to 
the quenched limit) and is shown in figure (\ref{fig:rho}).}
\label{tab:rhodyn}
\end{table}

\begin{figure}[hbt]
\hskip1in
\epsfysize=80mm
\epsffile{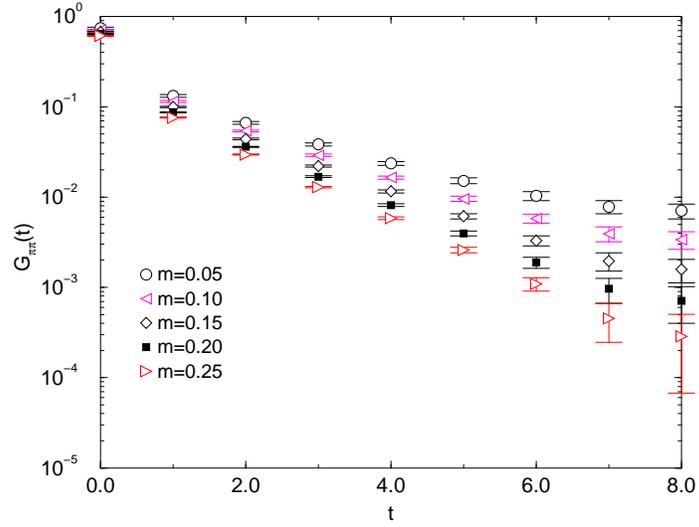}
\caption{\it This figure shows the zero momentum correlation 
function $G_{\pi\pi}(t)$ as a function of time slice.}
\label{fig:mescorr}
\end{figure}

\begin{figure}[hbt]
\hskip1in
\epsfysize=80mm
\epsffile{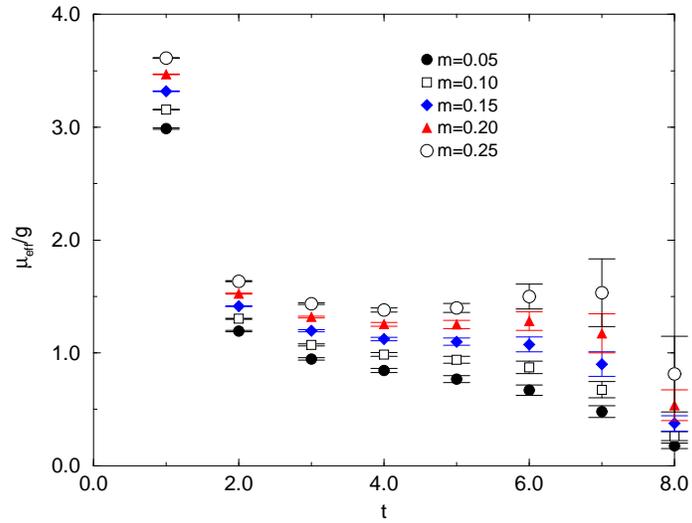}
\caption{\it The effective meson mass at various time slices is plotted
here to show evidence for plateaus.}
\label{fig:effm}
\end{figure}

\begin{figure}[hbt]
\hskip1in
\epsfysize=80mm
\epsffile{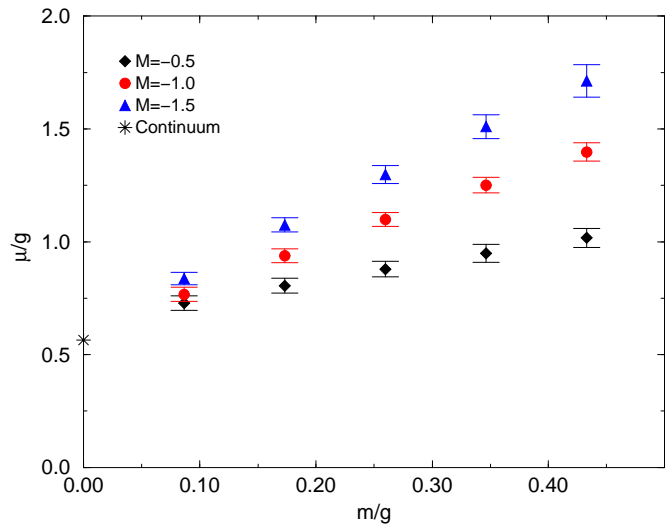}
\caption{\it The meson mass is plotted as a function of the fermion mass
for various values of $M$. The value in the continuum massless Schwinger 
model is represented by a star at $m/g=0$.}
\label{fig:meson}
\end{figure}

Next we look at the two point meson correlator as defined by
\begin{eqnarray}
G_{\pi\pi}(t) &=& \frac{1}{L_s}\sum_{x,y} 
\left\langle \left( {\rm tr}\left\{
\langle x,0|\gamma_5\frac{1}{D+m}|y,t\rangle
\langle y,t|\gamma_5\frac{1}{D+m}|x,0\rangle \right\}\right.\right.
\nonumber \\
& & - \left.\left.
{\rm tr}\left\{\langle x,0|\gamma_5\frac{1}{D+m}|x,0 \rangle\right\}
{\rm tr}\left\{\langle y,t|\gamma_5\frac{1}{D+m}|y,t \rangle\right\}
\right) [{\rm Det}(D+m)]^{N_f}\right\rangle
\nonumber \\
& & \;\;\;\;\;\;\;\;\;
/\left\langle [{\rm Det}(D+m)]^{N_f}\right\rangle
\end{eqnarray}
in order to extract the meson mass $\mu$. The kets $|x,t\rangle$ represent 
the normalized position basis states and the trace is over the spinor index.
Using the eigenvalues and eigenstates 
of $D$ it is easy to evaluate the propagators necessary for the calculation 
of $G_{\pi\pi}(t)$. In figure (\ref{fig:mescorr}) we plot it as a function of 
$t$. The function appears to fit to the usual form of $A\cosh(m(t-T/2))$, 
suggesting that conventional methods will be useful to extract the meson 
masses. A simple estimate of the meson mass can be made using the effective 
mass analysis. Defining the effective mass by
\begin{equation}
\mu_{\rm eff}(t) = -\log(G_{\pi\pi}(t)/G_{\pi\pi}(t-1)).
\end{equation}
we plot the results in figure (\ref{fig:effm}), which shows evidence for 
plateaux. We chose the values of $\mu_{\rm eff}$ at $t=5$ to estimate 
the meson mass. The meson masses are plotted in figure (\ref{fig:meson}) as 
a function of the fermion mass. The continuum result in the Schwinger model, 
$\mu/g = 1/\sqrt{\pi}$ is shown on the $m=0$ axis. Again our results
agree with in the expected $10-15\%$ error margin. It is interesting to
note that no factors of $Z$ are necessary here as this is a spectral 
quantity. However, the dependence of the meson mass on $M$ indicates a
renormalization of the fermion mass and the coupling. This effect can be 
studied more concretely and is left for the future.

\begin{center}
\noindent{\bf 6. Discussion and Conclusions}
\end{center}

  We have reproduced the essential features of the Schwinger model rather
easily on a coarse lattice using Neuberger's formulation of 
Ginsparg-Wilson fermions. This was possible because the physics of the 
one flavor model is essentially governed by the topological sectors, which 
was reproduced quite well on a moderately coarse and finite lattice.
We discovered an interesting connection between the zero modes and 
number of fermion flavors as a function of $M$. The value of $M$ was 
important in getting the right connection between the number of flavors 
and the zero modes of the Dirac operator. In the region $0 > M > -2$, 
where the theory is expected to describe the right physics the 
dominant effect of $M$ in the chiral condensate was a wave-function 
renormalization. Evidence for the renormalization of the fermion mass and
the coupling was also seen.

All this means that questions that are dominated by topological effects 
in QCD, such as the $\eta^\prime$ mass, may also be easily addressed with 
these fermions, once the right region in $M$ is isolated. Reproducing the 
physics of the anomaly has other interesting consequences.
In particular it can help in a better understanding of the QCD phase 
transition. Most calculations in lattice QCD indicate a second order 
phase transition for two flavor QCD. The most reliable calculations come 
from staggered fermions which have a remnant chiral symmetry. However, 
we have seen that staggered fermions produce very weak anomalous 
effects. These results have been confirmed in QCD just above the phase 
transition \cite{Christ}. As is well known \cite{Wilczek}, a very weak 
anomaly close to the transition can make the transition first order.
Since this is not seen in simulations, our understanding of the 
two flavor transition may not be complete. On the other hand, since 
the Ginsparg-Wilson fermions reproduce the effects of the anomaly rather 
easily and in 
addition have the correct flavor structure, they may give a more definitive 
understanding of the chiral transition. First such results have been
reported in the domain wall formulation \cite{PChen98}.
The present study also indicates that the Ginsparg-Wilson fermions 
behave quite differently with respect to quenching: another feature 
of reproducing the zero mode physics well. Hence the effects of the 
fermion determinant will play a more interesting role in QCD and 
quenched singularities \cite{Sharpe} could be easily seen with these
fermions. This is perhaps the easiest to study, since it involves only 
quenched calculations. 

One of the striking feature of QCD is the spontaneous breaking of chiral 
symmetry. Unfortunately this cannot be addressed in a two dimensional 
model due to the Mermin-Wagner theorem. On the other hand, there is very 
interesting physics associated with the many flavor Schwinger model.
For example one expects a non-analytic dependence of the meson mass and 
the chiral condensate on the fermion mass \cite{Coleman76}. As far as we know 
this has not been reproduced in the past mainly due to the lack of a good 
fermion formulation\footnote{This has been discussed for the first time
with domain wall fermions in Ref. \cite{Pavlos98}}. We plan to extend the 
present calculation in that direction. However, since we cannot expect 
non-analytic behavior to emerge on a finite lattice, a more careful finite 
size scaling analysis would be necessary. Thus the many flavor Schwinger 
model will
be considerably more difficult, but perhaps can serve as a useful guide 
for the study of chiral symmetry in QCD.

\begin{center}
{\bf Acknowledgment}
\end{center}
I would like to thank W. Bietenholz, T. Bhattacharya, T.-W. Chiu, N. Christ, 
R. Edwards, P. Hasenfratz, I. Hip, R. Mawhinney, R. Narayanan, 
F. Niedermayer, A. Smilga and P. Vranas for helpful discussions. 
As this article was about to be submitted I came to know of the work of 
T.-W. Chiu \cite{Chiu9810}, who has also analyzed the role of $M$ discussed 
in this article. I also thank W. Bietenholz for his suggestions after
a critical reading of the manuscript.

\newpage

\end{document}